\documentstyle[prb,twocolumn,aps]{revtex}

\begin{document}
\draft
\title{ Absence of Overscreened Kondo Effect in
Ferromagnetic Host}
\author{Jian-Hui Dai}
\address{
Zhejiang Institute of Modern Physics, Zhejiang University,
Hangzhou 310027, People's Republic of China}
\author{Yupeng Wang}
\address{Institute of Physics, Chinese academy of Sciences,
Beijing 100080,  People's Republic of China}

\maketitle

\begin{abstract}
We study the low temperature behavior of a boundary magnetic impurity
$S'=1/2$ in an open ferromagnetic Takhatajian-Babujian spin-$S$ chain.
For antiferromagnetic Kondo
coupling, it is shown via Bethe ansatz solution that the impurity spin is always locked into the critical behavior of the bulk.  At low temperature, a local composite of spin-($S-1/2$) forms near the impurity
site and its contribution to specific heat is of  simple power law 
$T^{\frac{1}{2}}$. The absence of overscreened Kondo effect is due to the
large correlation length of host spins which is divergent near the
quantum critical point.    
\end{abstract}


Multichannel Kondo problem was originally proposed to study the magnetic
impurity behavior in real metals\cite{nozieres}.  At low temperature for
antiferromagnetic Kondo coupling, the impurity spin $S'$ is completely
compensated by the surrounding $n$-channel electrons when
$2S'=n$; while it is undercompensated for $2S'>n$ or overcompensated for
$2S'<n$. The later situation (so-called overscreened Kondo effect) bears
no resemblance to a non-interacting gas of electrons\cite{anderi1} and provides us a
possible interpretation of non-Fermi liquid behavior observed in certain
uraniun alloys\cite{cox} as well as in certain tunneling problems in
two-level systems\cite{ralph}. As this effect is immediately destroyed by 
weak anisotropy, it has been a challenge to deduce its
signature relevant to other anomalous behavior in a variety of
exotic metals, especially when the systems show reduced effective
dimensionality, or undergo the proximity to  quantum critical phase
transitions\cite{andraka}. In this paper we will concentrate on multichannel Kondo problem in a typical quantum critical ferromagnet, with the number of channels $n$ being represented by the host spin, $n=2S$. Contrary to the  problem in antiferromagnetic host chain,  where overscreened Kondo effect appears when $S>S'$ ,  we will find that due to the divergent correlation length near the critical point, the overscreened Kondo effect does not exist in
ferromagnetic host.   

While the Kondo problems in paramagnetic or antiferromagnetic hosts
have been intensively investigated over three decades, they attract little
attentions in ferromagnetic host. The main reason is the difficulty 
in approaching this problem, i.e., usual  perturbation techniques as well as bosonization fail in the critical region because of the absence of the ``large energy scale", i.e., the Fermi energy. Recent investigations have been shown that as the system closes to a quantum critical point, the impurity behavior depends strongly on the host properties and  is non-universal\cite{cassanello}.
Another reason is due to the fact that the
existence of Kondo effect in ferromagnetic host remains controversial. 
At a first glance, the traditional Kondo screening seems possible only at temperature higher than the
Curie point where host magnetization vanishes. Otherwise, a local impurity
could not change the total magnetization, opposite to what happened in
dilute alloys. However, as far as the screening of a local moment is concerned, the issue should be how the impurity behavior changes as the temperature lowers down. The above picture does not necessary hold true even for the host goes away from the critical point  where the long range order is destroyed by strong 
fluctuations. Yet there is no rigorous proof of the absence of overscreening in low temperature ferromagnet .
In a earlier publication, Larkin and Mel'nikov  first studied the Kondo effect in an almost ferromagnetic metal\cite{larkin}. With traditional perturbation
theory they shown that the impurity susceptibility is almost Curie type
with logarithmic corrections at intermediately low temperature and the usual Kondo screening is not effective. More
recently,  the authors studied a similar problem in 1D, i.e., a boundary Kondo impurity of spin $S'$ in
$S=1/2$ ferromagnet\cite{wangd}. The model provides us  a first exact soluble example for the Kondo problem near a quantum critical point. Via Bethe ansatz solution it has been shown that at sufficient
low temperature, the impurity spin is completely screened by $2S'$ host
spins. Thus a local composite forms near the impurity site and 
shows a simple power law dependence of specific heat on temperature. Therefore, these interest results strongly motivate us to know what will happen for a magnetic impurity in a multichannel ferromagnetic host at low temperature and if there exhibits the conventional overscreened Kondo effect.

In order to study the multichannel Kondo problem in ferromagnet via exact 
solution,  we consider a magnetic impurity $S'=1/2$ coupled to an open
spin-$S$ chain which is integrable by Bethe ansatz. A well known
integrable extension of isotropic $S=1/2$ spin chain to arbitrary spin $S$
is\cite{kulish,tb,haldane}  
\begin{eqnarray}
H_S=J_0\sum^{N-1}_{j=1}Q_{2S}(\vec{S}_j\cdot\vec{S}_{j+1}),
\label{sh}
\end{eqnarray}
where $Q_{2S}(x)$ is a polynomial of degree $2S$ of $SU(2)$ invariant 
quantities $x=\vec{S}_j\cdot\vec{S}_{j+1}$. The general construction for $Q_{S}(x)$ is given in Ref.[10,11],   
one recovers the standard Heisenberg model 
and Takhatajian-Babujian model for $S=1/2$ and $S=1$ respectively.
Here we follow the definition of $H_S$ appeared in Ref.[11] so that the single-magnon dispersion  coincides with
that of the standard Heisenberg ferromagnet. Thus for $J_0<0$, the model (\ref{sh}) is qualitatively
equivalent to the standard spin-$S$ ferromagnetic Heisenberg chain\cite{haldane}.    
  The boundary impurity spin is now coupled to the first host spin as
\begin{eqnarray}
H_{imp}=J\vec{S}_1\cdot\vec{S'}.
\label{imph}
\end{eqnarray} 
The novelty of this construction is that the total
Hamiltonian $H_S+H_{imp}$ is again integrable for any $J_0$ and
$J$\cite{wang,daiw}.
The result is compared with the problem of a single impurity in periodic
spin chains\cite{anderi2}, where the integrability requires a fixed Kondo
coupling  $J$ and a fine turned impurity interaction instead of
simple form (\ref{imph}). As long as 
spin dynamics is concerned, the integrable spin-$S$ chain represents the 
multichannel host, because Kondo screening of the impurity is
equivalent to usual multichannel Kondo problem in antiferromagnetic
case($J_0>0$)\cite{anderi2}: the impurity spin is 
completely screened for $S'=S$; partially screened for $S'>S$, with
Schottky anomaly when an external magnetic field $H$ is applied; and
overscreened for   $S'<S$, giving rise to quantum critical
behavior.

The model Hamiltonian is solved within the framework of Bethe ansatz approach to
impurity problem\cite{daiw,anderi3}. Eigenstates of the model are
parameterized by a set of rapidities $\{\lambda_{\alpha}\}_{\alpha=1}^{M}$  
($M$ being the number of down spins), with  total magnetization
$S_z=NS+1/2-M$. Each rapidity represents
a magnon of energy $-J_0 S/(\lambda_{\alpha}^2+S^2)$, they are solutions of 
Bethe ansatz equations 
\begin{eqnarray}
e_{imp}(\lambda_{\alpha})e_{2S}^{2N}(\lambda_{\alpha})
=\prod_{\pm}\prod_{\beta\neq\alpha}^M e_{2}(\lambda_{\alpha}\pm \lambda_{\beta})
\label{bae}
\end{eqnarray}
where $e_r(x)=\frac{x+ir/2}{x-ir/2}$, $e_{imp}(x)=e_{1+2c}(x)e_{1-2c}(x)$
with $c=\sqrt{(S+1/2)^2-J_0/J}$. In the following, we only consider the case $J_0<0, J>0$, so $c\geq S+1/2$. In the thermodynamics limit the
solutions are  classified by $n$-strings or  $n$-magnons, they are associated with functions   
$\eta_n(\lambda)$ ( with real rapidity $\lambda$ \cite{note1}), which in
 thermal equilibrium for finite temperature $T$ and field $H$ satisfy 
an infinite set of non-linearly coupled integral equations:
\begin{eqnarray}  
\ln \eta_n
=\frac{\pi}{g}\frac{|J_0|}{T}\delta_{n,2S}
+ {\bf G}\ln[(1+\eta_{n-1})(1+\eta_{n+1})],
\label{inte2}
\end{eqnarray}
with the asymptotic condition $\lim_{n\to\infty}\frac{\ln\eta_n(\lambda)}n=H/T\equiv 2x_0$.
Here, ${\bf G}$ 
is an integral operator with kernel $g(\lambda)=1/2\cosh(\pi\lambda)$. 
 Notice that the free energy, expressed in terms of $\eta_n(\lambda)$, consists of three parts
$F=F_{bulk}+\frac{1}{N}(F_{imp}+F_{edge})$, 
bulk part is always the dominate one, impurity and edge parts are both of  
order of $1/N$, edge part is 
only due to the open boundary condition and is irrelevant to spin dynamics. 
We will discuss bulk and impurity parts in the following.

\par
{\sl Ferromagnetic host}.
 The groundstate of ferromagnetic host has $M=0$ with finite  
magnetization $S$ per site. The pure states are those with spin aligned
along an arbitrary axis. 
With finite $T$ and $H$ the elementary excitations are $n$-magnons, bound states
carrying $n$ quanta of reduction of magnetization along the ordered axis. Each
$n$-magnon has energy $\epsilon_n=2|J_0|a_{n,2S}(\lambda)$ with crystal momentum $K=4\sum_{p=1}^{n,2S}\arctan \frac{n+2S+1-2p}{2|\lambda|}$. In open boundary
problem, $K$ is positive. The energy spectrum $\epsilon_n(K)$
is a continuous function, increases monotonically as $K$ varies in $[0,\pi
min(n,2S)]$. The long-wavelength limit, i.e., $K\sim 0$,  
$\epsilon_n(K)\sim |J_0|S
K^2/n$, corresponds to $|\lambda|\sim\infty$. The absence of gap for all
$n$-magnons at boundaries of Brillouin zones $\pi min(n,2S)$   indicates
the exact
cancellation of Umklapp processes\cite{haldane}. Correspondingly in thermodynamics limit,  the driving term  in Eq.(\ref{inte2}) dominates when
$T\rightarrow 0$,  all $\eta_n(\lambda)$ functions are
relevant.
The situation is quite different to antiferromagnetic host ($J_0>0$), where only
$\eta_{2S}(\lambda)$ is relevant as $T\rightarrow 0 $. So the solutions
of Eq.(\ref{inte2}) are more complicated for $J_0<0$ . Here two limiting cases are
important: (a) Weak coupling limit:
$T\rightarrow\infty$. The driving term disappears, there is only one 
parameter in equations, $x_0=H/T$, the solutions are 
\begin{eqnarray}
\eta_n=\frac{\sinh^2[(n+1)x_0]}{\sinh^2x_0}-1
\label{weak}
\end{eqnarray}
for all $\lambda$ and $n$, this is the free spin limit.      
(b) Strong coupling limit: $T\rightarrow 0$. In this case the driving term
diverges, $\ln(1+\eta_n)\sim \ln\eta_n$, we obtain
\begin{eqnarray}
\eta_n(\lambda)=\exp[(
nH+\pi a_{n,2S}(\lambda)|J_0|)/T].
\label{strong}
\end{eqnarray}
At zero temperature, $\eta_n=\infty$ for all $n$, i.e., no magnons,  the
ferromagnetic groundstate with $S_z/N=S$ and $E/N=H$ is reproduced.
At low but finite temperature, $\eta_n(\lambda)$ as function of
$\lambda$ and $n$ show crossovers between the two asymptotic solutions (\ref{weak}) and
(\ref{strong}). In principle, they can be solved numerically for fixed $T$ and $H$.  To study the critical behavior of $S=1/2$ ferromagnetic Heisenberg chain,
Schlottmann\cite{schlottmann1}
proposed an analytic method  to solve these equations, based on an
elegant but simple correlation-length approximation, and the result  
coincides with the numerical one very well\cite{schlottmann1,schlottmann2}. His idea is now
developed for arbitrary $S$ as follows.

Let us first notice that for sufficient large $|\lambda|$, the driving term  
becomes negligible and one reproduces the weak coupling solution Eq.(\ref{weak})
for all $n$, up to 
${\cal O}(e^{-\pi |\lambda|}|J_0|/T)$.
Similarly, for sufficient large string index $n$, $\eta_n(\lambda)$ is
again  close to Eq.(\ref{weak}), because the effect of the driving term is also
negligible. On the other
hand, for small $|\lambda|$ and $n$,
the driving term dominates, $\eta_n(\lambda)$ approach the strong coupling  
solution (\ref{strong}). For intermediate $\lambda$ and $n$ we have a crossover region. 
By equating two solutions (\ref{weak}),(\ref{strong}) for small $\lambda$ and $H$, we define a
crossover scale $n_c(T)$
as 
$2S|J_0|n_c=T\ln[n_c(n_c+1)]/2$,
or solving it iteratively for $T<<|J_0|$,
\begin{eqnarray}
n_c(T) \approx 
\frac{2S|J_0|}{T}
[\frac{1}{\ln(\frac{2S|J_0|}{T})}+\frac{\ln\ln(\frac{2S|J_0|}{T})
}{\ln^2(\frac{2S|J_0|}{T})}+\cdots]. 
\end{eqnarray}
The meaning of $n_c(T)$ is clear: for $n > n_c$, $\eta_n$ is close
to weak coupling solution while for $n < n_c$ it is close to strong
coupling  one.  Therefore $n_c(T)$ serves as the  correlation
length of host, because it
is the average number of the correlated spins in thermal equilibrium.
Similarly, for each $n$-magnon, we define a crossover scale for rapidity,
\begin{eqnarray}
\lambda_{n}^c(T) \approx \sqrt{\frac{n}{\ln(n+1)}\frac{2S|J_0|}{T}}~.
\end{eqnarray}
Obviously $\lambda_{n}^c(T)$ is the correlation length in momentum space: for
$|\lambda|>\lambda_{n}^c(T)$ (long wave-length or low energy),
$\eta_n(\lambda)$ is close to weak coupling solution,
while for $|\lambda|<\lambda_{n}^c(T)$ (and $n < n_c$) it is close to strong 
coupling one. 
Thus we adopt the following strategy to solve $\eta_n(\lambda)$, or more conveniently, to calculate $\zeta_n(\lambda)=\ln[1+\eta_n(\lambda)]-[\pi a_{n,2S}(\lambda)|J_0|+nH]/T$: 
(i) As a zero order approximation, we assume 
$\eta_n(\lambda)=\infty $ for $ n < n_c$ and $|\lambda|<\lambda_{n}^c$ , and
elsewhere is given by Eq.(\ref{weak}).
(ii) Substituting this approximation into the right hand side
of Eq.(\ref{inte2}), we get the first order approximation for $\zeta_n(\lambda)$.
The result is 
\begin{eqnarray}
\zeta_n(\lambda)\approx  
\sum_{m=1}^{n_c}\{\ln[1+\frac{1}{m(m+2)}]-\frac{2}{3}x_0^2\} 
 B_{mn}(\lambda_m^c-\lambda)\\ \nonumber
+2n \ln \frac{\sinh (1+n_c)x_0}{\sinh n_c x_0}, 
\label{zeta}
\end{eqnarray}
with $B_{mn}(\lambda_m^c-\lambda)=(\int_{-\infty}^{-\lambda_m^c}+
\int^{\infty}_{\lambda_m^c})A_{mn}(\lambda-\lambda')d\lambda'$.
The leading contribution to bulk free energy $F_{bulk}=-T\int d\lambda g(\lambda) \zeta_{2S}(\lambda)$ for small  $T$ and $H/T$ 
is estimated as 
\begin{eqnarray}
F_{bulk}=-1.1 (\frac{2S}{|J_0|})^{1/2}T^{3/2}~~~~~~~~~~~~~~~~~~~~~~~~~ \nonumber\\
-0.42|J_0|\frac{(2SH)^2}{T^2}[\frac{1}{\ln(\frac{2S|J_0|}{T})}+
\frac{\ln\ln(\frac{2S|J_0|}{T})}{\ln^2(\frac{2S|J_0|}{T})}+\cdots]
\label{hoste}
\end{eqnarray}
The zero field dependence of the free energy, as given above, is due to the $n$-magnon contributions from those within the cut-off, $n_c$; while the contributions beyond the cut-off is of higher order, $\sim T^2$. So this part eventually approach the exact one as $n_c\rightarrow\infty$ or $T\rightarrow 0$.
The finite-field contribution of the free energy, which  comes from all kinds of $n-$magnons, is a result of competition of both parts, and it is approximately proportional to the correlation length $n_c$, giving rise logarithmic corrections to the bulk susceptibility. Notice that by rescaling $2SJ_0 \rightarrow J_0$, the bulk free energy is proportional to that of the standard spin$-1/2$ ferromagnetic Heisenberg chain, so one obtains $C_{bulk}=2SC_{1/2}\sim T^{1/2}$ and $\chi_{bulk}=2S\chi_{1/2}\sim T^{-2}\ln^{-1}T$.

\par
{\sl Boundary impurity}.
When the impurity is coupled to the ferromagnetic host via antiferromagnetic $J>0$, Eq.(\ref{bae}) has a pure imaginary mode $\lambda=i(c-1/2)$. This mode contributes a negative energy
$\varepsilon_{imp}=|J_0|S/[S^2-(c-1/2)^2]<0$ to the system, indicating the formation of a bound state, i.e., a local composite made of the impurity and its neighboring spin S. One notices that $\lambda=i(c-1/2)$ is the only possible boundary mode for $S'=1/2$ and no boundary string\cite{wangd} is allowed
because $\lambda=i(c+1/2)$ is forbidden as shown in Eq.(3).

Therefore, by taking into account the pure imaginary mode, the BAE  
for the bulk rapidities read
\begin{eqnarray}
e_{2S}^{2N}(\lambda_{\alpha})
\frac{e_{2c-3}(\lambda_{\alpha})}{e_{2c-1}(\lambda_{\alpha})}=\prod_{\pm}\prod_{\beta\neq\alpha}^{(M-1)}e_{1}(\lambda_{\alpha}\pm \lambda_{\beta}).
\label{bbae}
\end{eqnarray}
When $2c$ is an integer, the impurity free energy is the difference of contribution from two effective "ghost" spins $S_+=c-1/2$ and $S_-=c-3/2$~\cite{daiw}
\begin{eqnarray}
F_{imp}=-\frac{T}{2}\int d\lambda g(\lambda)[\zeta_{2c-1}(\lambda) -\zeta_{2c-3}(\lambda)].
\label{impf}
\end{eqnarray}
One finds that the leading contributions from the impurity is the same as that of $\frac{1}{2}-$spin(rescaling of $J_0$ is used), but is always negative, i.e., $C_{imp}=-C_{1/2}
\sim T^{1/2}$, $\chi_{imp}=-\chi_{1/2}\sim (T^2\ln T)^{-1}$, indicating the freezing of some
degrees of freedom of the bulk. 
Notice that $c$-dependence arises only in the next orders. 
When $2c$ is not an integer, the situation is somewhat more complicated. In this case, there are additional contributions coming from the non-zero residue $\{2c\}=2c-[2c]$ ($[2c]$ is the integer part of $2c$), which renormalize the effective Kondo temperature $T_k\sim 1/\cos(\pi\{2c\}/2)$ as well as the physical quantities. Explicit calculation shows that they only change the subleading behavior.  

The effect of the local composite can be analyzed in the same way. This local composite is made of the impurity spin and the neighboring host spin. In contrast to the single channel problem where the local composite is always a spin singlet, the present one is not a spin singlet, but with a residual spin $S-1/2$. Its feature is encoded in BAE as follows 
\begin{eqnarray}
e_{S}^{2(N-1)}(\lambda_{\alpha})
=\exp \{i\phi(\lambda)\}\prod_{\pm}\prod_{\beta\neq\alpha}^{(M-1)}
e_{1}(\lambda_{\alpha}\pm \lambda_{\beta}).
\label{locbae}
\end{eqnarray}
where   
$\phi(\lambda)=-i\ln[e_{2S}^{2}(\lambda)\frac{e_{2c-3}(\lambda)}{e_{2c-1}
(\lambda)}]
$
represents the phase shift of a bulk spin wave scattering off the local composite. In the limit $J\rightarrow \infty$, one finds a momentum dependent phase shift $\phi(\lambda) \neq 0$ indicating no complete compensation of the impurity spin in multichannel ferromagnetic host. (except for $S=1/2$ where $\phi(\lambda) = 0$). Similarly, For $2c=$integer, we estimate the local composite contribution to the free energy
\begin{eqnarray}
F_{loc}=\frac{T}{2}\int d\lambda g(\lambda)[ 2\zeta_{2S}(\lambda)+\zeta_{2c-3}(\lambda) -\zeta_{2c-1}(\lambda)].
\label{locf}
\end{eqnarray}
One immediately obtains $C_{loc}=(S-1/2)C_{1/2}\sim T^{1/2}$, $\chi_{loc}=(S-1/2)\chi_{1/2}\sim (T^2\ln T)^{-1}$.
So the leading order behavior of the local composite is similar to that of ($S-1/2$)-spin, 
indicating that the impurity is neither completely screened as in one channel problem nor overscreened as in antiferromagnetic host, but is always locked into the critical behavior of 
the bulk.  

\par
{\sl Discussions and conclusions}. 
In this paper, we studied a spin-1/2 boundary impurity coupled to the ferromagnetic spin-$S$ chain. The  results were obtained with the assumption that the crossover from strong coupling to
 weak one is abrupt. A smooth crossover region should not modify the low-$T$ and small $H/T$ dependences of the free energies obtained. Moreover,  if the correlation lengthens $n_c$ and $\lambda^c_n$ are scaled, the low-$T$ dependence of our results remain
unchanged,  only the amplitudes are rescaled. The leading zero-field dependence of the bulk free energy is a rigorous result, even the amplitude agrees well with those obtained numerically for $S=1/2$~\cite{note}, or analytically for $S>1/2$ via  spin-wave theory\cite{fisher}. However, since all $n-$strings contribute, the situation is more complicated for the susceptibility, logarithmic corrections in susceptibility arise even without Kondo impurity, indicating the existence of a marginal variable which does not exist in classical spin wave solution.     

The impurity effects are controlled by the bulk properties and the impurity coupling. In the case of antiferromagnetic host, see Ref.[12,13], a boundary impurity will shows the usual Kondo effect when the impurity spin $S'$ is less than or equal to the host spin $S$, i.e., $S'>S$, otherwise, it shows a typical overscreened Kondo effect, i.e., $S>S'$. Depending on the Kondo coupling $J$, or $c$, there exhibits some novel quantum critical phenomenon in both cases. In the case of ferromagnetic host, the single channel problem ($S=1/2, S' \geq 1/2$) shows a novel Kondo diamagnetic effect, the impurity spin is completely screened by surrounding $2S'$ host spins for antiferromagnetic $J>0$. Other magnons see a local composite formed around the impurity site and interact with it weakly.  

The situation for multichannel ferromagnetic is dramatically changed. Due to the presence of antifrromagnetic coupled impurity, there is  always a pure imaginary mode in addition to the bulk $n$-magnons. The magnetization is suppressed due to the local composite which is a spin $S-1/2$ object. But the impurity spin is not overscreened, in contrast to that in antiferromagnetic host. Rather, it is locked into the critical behavior of the bulk, with impurity specific heat showing a simple power law $C_{imp}\sim T^{\frac{1}{2}}$. The local composite now effectively coupled {\it ferromagnetically} to the other host spins via quantum fluctuation, showing the same power law  specific heat $C_{loc}\sim T^{\frac{1}{2}}$. The absence of multichannel Kondo effect is due to the large correlation length $n_c$ which is divergent when $T\rightarrow 0$. Though these results are limited to 1D , similar phenomena may exist in higher dimensions. In multichannel ferromagnet, the antiferromagnetic Kondo coupling always leads to a local composite of a smaller spin, and leads to ferrimagnetic state when there is a finite density distribution of the impurities.

\vspace{5mm}
\par 
The authors are grateful to Profs. N. Anderi, I.N. Karnaukhov and Yu Lu for helpful discussions. We acknowledge the
financial support of NSF of China.


\begin{references}
\bibitem{note1}More precisely, $\lambda$ in $\eta_n(\lambda)$ or $\zeta_n(\lambda)$ defined later could be complex, but limited to the condition $|Im \lambda|<1/2$.
\bibitem{note}Numerical bulk results for $S=1/2$ are obtained by many people, see References in Ref.[18]. For ferromagnetic chain, the results are expected to be valid for $S>1/2$, up to a numerical factor dependent on $S$.
\end{references}
\end{document}